\documentclass[conference]{IEEEtran}
\IEEEoverridecommandlockouts
\usepackage{cite}
\usepackage{amsmath,amssymb,amsfonts}
\usepackage{algorithmic}
\usepackage{graphicx}
\usepackage{textcomp}
\usepackage{xcolor}
\def\BibTeX{{\rm B\kern-.05em{\sc i\kern-.025em b}\kern-.08em
    T\kern-.1667em\lower.7ex\hbox{E}\kern-.125emX}}
\begin{document}

\title{Blocker-Aware Beamforming and Dynamic Power Allocation for Multicarrier ISAC-NOMA Systems}
\author{
	\IEEEauthorblockN{Abdulahi Abiodun Badrudeen, Nakyung Lee, Adam Dubs, Sunwoo~Kim}
	\IEEEauthorblockA{Department of Electronic Engineering, Hanyang University, South Korea
		\\\{aabadrudeen, nagyeong2379, duanandi16, remero\}@hanyang.ac.kr}}
\maketitle

\begin{abstract}
This paper proposes a blocker-aware multicarrier integrated sensing and communication (ISAC)-non orthogonal multiple access (NOMA) system, leveraging hybrid beamforming and dynamic power allocation to enhance spectrum efficiency in 6G networks. Recognizing the performance degradation caused by environmental blockers, the system introduces a joint waveform design that ensures robust operation under varying channel conditions. A channel switching mechanism is deployed to reroute communication through alternative non-line-of-sight paths when the primary line-of-sight links are obstructed. Moreover, a dynamic power allocation strategy enforces a minimum rate constraint for the weak NOMA user, ensuring consistent quality of service. Extensive simulations over multiple blockage scenarios and signal to noise (SNR) conditions validate the effectiveness of the proposed solution. Notably, under severe blockage, the system achieves up to a 400\% sensing rate enhancement at 15 dB SNR, with only a 20\% reduction in communication rate. These results corroborate the system’s ability to adapt and optimize joint sensing-communication performance in practical deployment environments.
\end{abstract}


\section{Introduction}
A key component of 6G research is integrated sensing and communication (ISAC), which unifies sensing and communication operations using shared hardware and spectrum resources. Cutting-edge applications, namely autonomous vehicles, augmented reality, robotics, and simultaneous localization and mapping increasingly demand ISAC. These applications can be efficiently aided by ISAC networks, which leverage high-resolution sensing and ultra-wide bandwidth capabilities in the millimeter-wave and terahertz spectrum. However, these spectra are vulnerable to path loss fading and multipath effects. Massive  multiple input multiple output (MIMO) and hybrid beamforming (HBF) technologies have been established by researchers as part of the solutions to mitigate these effects, thereby ensuring a higher throughput for ISAC systems.

Hybrid (digital and analog) beamforming in multi-user massive MIMO systems, each user is limited to one beam via one radio frequency (RF) chain. Non-orthogonal multiple access (NOMA) scheme has been established as a solution to this hardware constraint of conventional multi-user MIMO system. NOMA allows more than one user to share the same beam and RF chain per symbol stream~\cite{1}. NOMA domains are categorized as power and code domains. In this paper, we adopt the power domain NOMA approach to study ISAC systems. Basically, power domain NOMA involves the superposition of users' symbols at the base station (BS) and successive interference cancellation at the users' end.

To achieve effective NOMA communication, user pairing and ordering must be performed at the BS before superposition coding of user's symbol streams in downlink scenario. In light of this, user's with highly correlated channels or in the same angle of arrivals (AoA) are paired to share the same beam. Leveraging this NOMA strategy to jointly transmit data symbol and pilot symbol has been studied in \cite{2,3,4}, thereby leading to NOMA-enabled ISAC and ISAC-enabled NOMA research. One of the issues that may hinder the performance of ISAC-NOMA is the sudden presence of static or mobile objects along the sensing or communication path. Detecting this blockage for ISAC-NOMA system to adaptively beamform and adjust power is critical for future generation networks.

\subsection{Related Studies}
Theoretical trade-offs between waveform design and sensing accuracy, as well as Cramér-Rao bound performance of ISAC were established by early foundational studies such as \cite{ 5} and \cite{6}. These studies offer important insights; however, they ignore dynamic interference scenarios and the chance for multi-user spectral efficiency improvements in favor of standalone ISAC systems. Although \cite{7} later attempts, matching ISAC criteria to 6G key performance indicators such as angular resolution and update rate, real-world issues like hardware flaws and mobile blockers were not addressed.

Integrating ISAC with NOMA has been investigated to meet spectral efficiency requirements. Early studies that examined NOMA-radar coexistence, including \cite{2}, found trade-offs between radar ambiguity and block error rate. Similarly, \cite{3} optimized NOMA for multicast-unicast communication and cooperative radar sensing. However, the potential of HBF and multicarrier (MC) systems to reduce inter-carrier interference (ICI) in wideband mmWave environments is ignored in current studies, which assume simplified beamforming designs and narrowband channels.

ISAC-NOMA integration is further made possible by recent developments in beamforming and MC systems. \cite{8}, for example, optimized mmWave channel estimation duration via adaptive pilot allocation, a technique that works for MC-HBF but hasn't been applied to ISAC-NOMA yet. While codebook-based beamforming methods in \cite{9} increase resilience to localization mistakes, they are unable to handle dynamic resource allocation or blocker-induced interference. In the meantime, joint waveform-phase shift optimization is demonstrated by RIS-assisted ISAC designs \cite{10}, which encourage MC-HBF-NOMA co-design but lack blocker-aware techniques.

A number of significant holes still need to be filled in spite of these developments. Firstly, there are a few studies that use ISAC's sensing capabilities, namely receive signal power from backscattered signal at the BS to detect blockers in real time and reallocate NOMA resources adaptively. Finally, current efforts, such as those cited in \cite{2,3,10}, do not adequately optimize HBF in conjunction with MC waveforms, which is crucial for reducing ICI and balancing hardware complexity. Hence, this research proposes a blocker-aware MC-HBF-ISAC-NOMA framework to overcome these deficiencies. The proposed system integrates high-resolution sensing and dynamic power domain NOMA resource allocation, to enhance spectral efficiency and resilience in 6G networks.

\section{System Model Presentation}

\begin{figure}[!t]
	\centering
	\includegraphics[width=1\columnwidth]{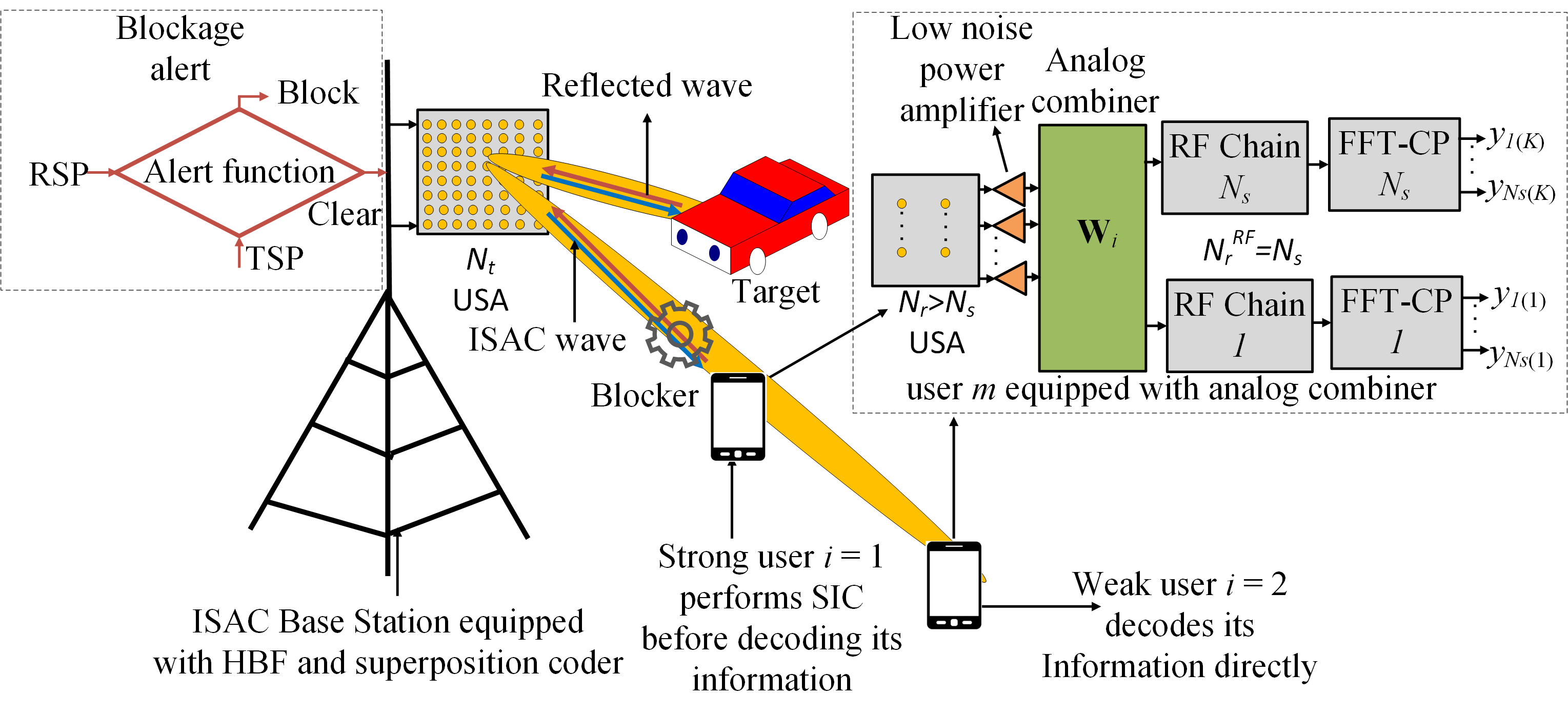}
	
	\caption{Illustration of an ISAC-enabled Blocker aware mmWave NOMA Communication system.}\label{Figure 1}
\end{figure}

We consider an ISAC system framework illustrated in Fig.~\ref{Figure 1}, where an ISAC BS equipped with a uniform square antenna array (USA) $N_t$ antennas, simultaneously performs downlink multi-user communication and monostatic radar sensing. Explicitly, best $N_R$ antennas are selected from $N_t$ BS antennas to receive the reflected sensing signals at the BS for rank sufficiency. In light of this, the BS serves two distinct functions using a joint HBF waveform, namely sensing a dedicated passive target using a sensing beam (beam 1), and communicating NOMA users sensing, exploiting the communication beam (beam 2), which also senses both user-side partial or temporary blockages owing to temporary presence of blocker like human body along the communication path. 

\begin{figure}[!t]
	\centering
	\includegraphics[width=1\columnwidth]{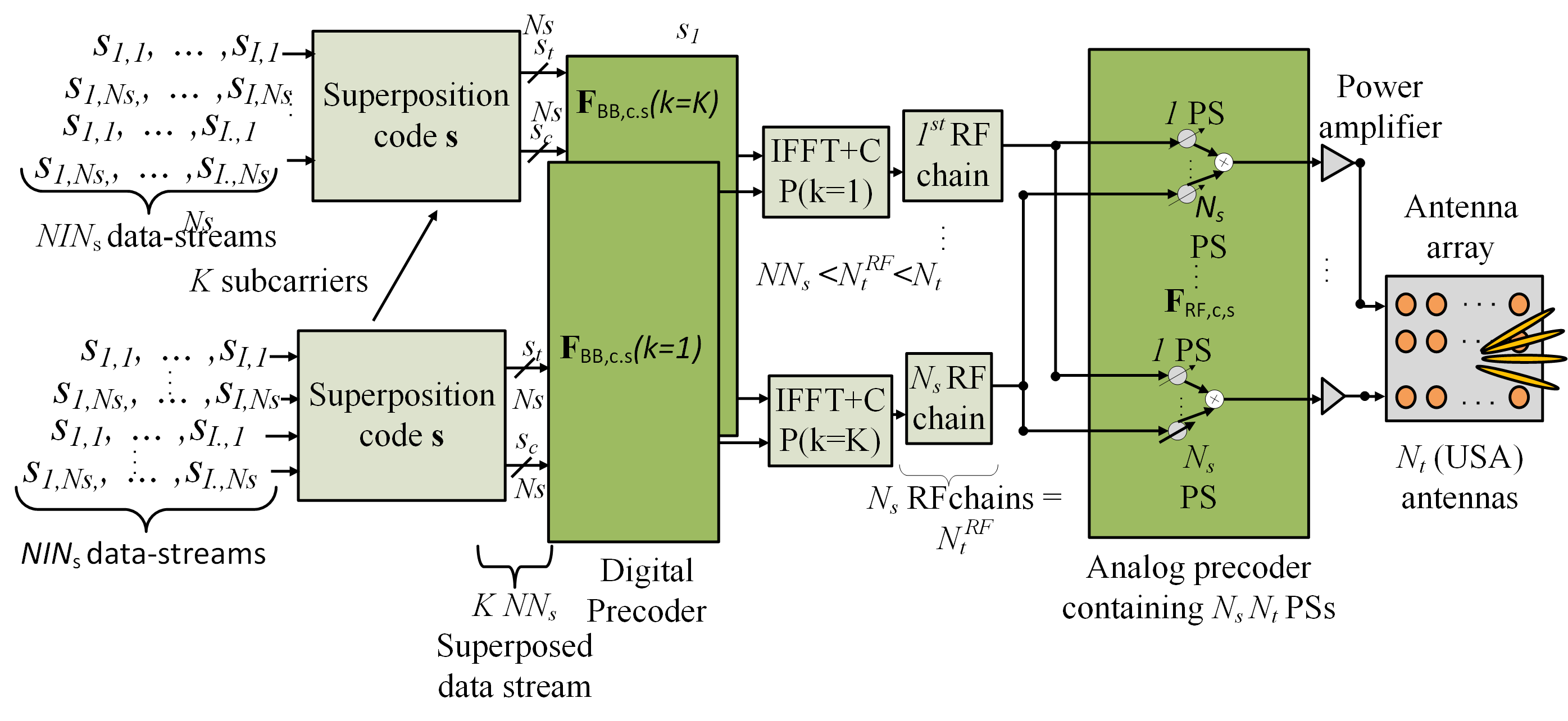}
	
	\caption{Superpositioning coding for HBF structure deployed at the BS section of Fig. \ref{Figure 1}. }
	\label{Figure 2}
\end{figure}

The proposed system operates over $K$ subcarriers in a MC framework illustrated in Fig.~\ref{Figure 2}, with hybrid analog–digital precoding employed at the BS. The joint communication and sensing precoding matrix on subcarrier $k$ is denoted as $\mathbf{F}_{c,s}(k) = \mathbf{F}_{\text{RF},{c,s}} \mathbf{F}_{\text{BB},{c,s}}(k)$, where $\mathbf{F}_{\text{RF},{c,s}} \in \mathbb{C}^{N_t \times N_t^{\text{RF}}}$ is the analog RF precoder and $\mathbf{F}_{\text{BB},{c,s}}(k) \in \mathbb{C}^{N_t^{\text{RF}} \times NN_s}$ is the baseband digital precoder for $NN_s$ superposed joint data and pilot (sensing) symbol streams, where $N_t^{\text{RF}}(=NN_s)$ denotes number of RF chains at the BS, $N$ and $N_s$ are the total number of clusters and transmit user's or sensing symbol streams, respectively. Thus, the joint ISAC transmit waveform can be modeled as $\mathbf{x}_{s,c}(k) = \mathbf{H}_{c,s}(k)\mathbf{F}_{\text{RF},{c,s}} \mathbf{F}_{\text{BB},{c,s}}(k)\mathbf{s}_{s,c}(k)$, where $\mathbf{H}_{c,s}(k)$ and $\mathbf{s}_{s,c}(k)$ denote ISAC channel matrix and pilot with data superposed symbols vector, respectively.

Each NOMA cluster contains a pair of users, namely a strong user and a weak user, sharing the same AoA and hence served using power-domain superposition coding. The users are equipped with $N_r$ USA antennas each and apply analog combining matrix $\mathbf{W}_i(k) \in \mathbb{C}^{N_r \times N_r^{\text{RF}}}$ for downlink decoding on subcarrier $k$, where $N_r^{\text{RF}}$ represents the number of RF chains at each NOMA user. A second beam, orthogonal to the NOMA users’ spatial domain, is reserved for probing a dedicated passive sensing target, such as a ground vehicle in this scenario, using the same superposed symbols deployed for communication. This prevents the sensing signal from interfering with the communication signal~\cite{4}. 

The proposed model distinguishes itself from prior ISAC works, such as \cite{4}, where different beams served NOMA users, by utilizing a single beam waveform for NOMA communication, and ditto for sensing, enabling two-dimensional beam control via the USA, supporting HBF-enabled MIMO-NOMA communication, and incorporating a blocker-aware architecture. The proposed ISAC-NOMA is limited to two clusters for the sake of clarity of analysis. However, the framework can be extended to more than two clusters and NOMA users. The goal is not to optimize sensing efficiency alone, but rather to enhance communication reliability by proactively detecting and mitigating blockage conditions through passive sensing.

\subsection{Communication Model}

The downlink communication signal is a superposition of NOMA-coded data streams:
\begin{equation}
	\mathbf{x}_{c}(k) = \mathbf{F}_c(k) \left( \sum_{i=1}^{2} \sqrt{p_i(k)} \mathbf{I}_{N_s} \mathbf{s}_i(k) \right),
\end{equation}
where $ \mathbf{F}_c(k)= \mathbf{F}_{\text{BB},c}(k)\mathbf{F}_{\text{RF},{c,s}}$ denote the hybrid precoding for communication on $k$-th subcarrier, $\mathbf{s}_i(k)$ is the data streams symbol for user $i$ on subcarrier $k$ and $p_i(k)$ is the power allocated and $\mathbf{I}_{N_s}$ is an identity matrix of an $N_s$ order. Without loss of generality, user 2 is the weak user and is assigned more power, namely, $p_2(k) > p_1(k)$.

The received signal at user $i$ on subcarrier $k$ is:
\begin{equation}
	\mathbf{y}_i(k) = \mathbf{W}_i^H(k) \mathbf{H}_i(k) \mathbf{x}(k) + \mathbf{W}_i(k)\mathbf{n}_i(k),
\end{equation}
where $\mathbf{H}_i(k) \in \mathbb{C}^{N_r \times N_t}$ is the downlink channel matrix from the BS to user $i$, and $\mathbf{n}_i(k)$ is zero-mean AWGN with variance $\sigma^2$ for $k=1, \cdots, K$ . The achievable rate for each user depends on standard NOMA decoding:
\begin{equation}
	R_i(k) = \log_2(\text{det}\left(\mathbf{I}_{N_s} + \text{SINR}_i(k) \right)),
\end{equation}
where $\text{det}$ indicates determinat, with $\text{SINR}_i(k)$ formulated based on perfect successive interference cancellation (SIC) at the strong user and using the same data symbols for both sensing and communication. That will ensure interference from sensing does not affect the NOMA users and vice versa~\cite{11}. Hence, the generalized SINR for each user $i = 1, \dots, I$ in the NOMA cluster is expressed as:

\begin{equation}
	\text{SINR}_i(k) = 
	\frac{ p_i(k) \left| \mathbf{W}_i^H(k) \mathbf{H}_i(k) \mathbf{F}_c(k) \right|^2 }
	{ \sum_{j > i} p_j(k) \left| \mathbf{W}_i^H(k) \mathbf{H}_i(k) \mathbf{F}_c(k) \right|^2 + \hat{\sigma}_i^2 },
	\label{eq:SINR_general}
\end{equation}
where $\hat{\sigma}_i^2 = \left| \mathbf{W}_i^H(k) \sigma \right|^2$ represents the effective post-combining noise power at user $i$.

Moreover, the total communication rate $ R_{\text{c-sum}}$ for all users over all subcarriers is obtained from:
\begin{equation}
	R_{\text{c-sum}} = \sum_{k \in \mathcal{K}} \sum_{i = 1}^{I} \log_2(\text{det}\left(\mathbf{I}_{N_s} + \text{SINR}_i(k) \right)),
	\label{eq:R_com_sum}
\end{equation}
where $\mathcal{K}$ is is the set of all subcarriers.

\subsection{Sensing Model}

The BS performs monostatic sensing by exploiting echoes received from the same waveform used for communication. These echoes originate from two types of reflectors: the user cluster (used to detect dynamic blockers) and a dedicated passive sensing target located in a separate angular domain. The received sensing signal on subcarrier $k$ at the BS is:
\begin{equation}
	\mathbf{y}_R(k) = \sum_{o \in \mathcal{O}} \rho_o(k) \mathbf{H}_o^H(k) \mathbf{F}_{s,o}(k) s_s(k) + \mathbf{n}_R(k),
\end{equation}
where $\mathcal{O} = \{\text{User 1}, \text{User 2}, \text{Target}\}$ is the set of reflectors, $\rho_o(k)$ denotes the reflection coefficient of object $o$, and  $ \mathbf{F}_{s,o}(k)= \mathbf{F}_{\text{BB},{s,o}}(k)\mathbf{F}_{\text{RF},{c,s}}$ denote the hybrid precoding for sensing each object on $k$ subcarrier. The receive combining vector at the BS is $\mathbf{W}_R(k)$.

The signal-to-interference-plus-noise ratio (SINR) for sensing reflection from object $o$ is given by:
\begin{equation}
	\text{SINR}_{o}(k) = 
	\frac{ p_o \left| \rho_o(k) \mathbf{W}_R^H(k) \mathbf{H}_o(k) \mathbf{F}_s(k) \right|^2 }
	{ \sum_{j \in \mathcal{O},\, j \neq o} p_j \left| \rho_j(k) \mathbf{W}_R^H(k) \mathbf{H}_j(k) \mathbf{F}_s(k) \right|^2 + \hat{\sigma}_R^2 },
	\label{eq:SINR_sense}
\end{equation}
and the corresponding sensing rate $R_{\text{s},o}(k)$ for subcarrier $k$ is expressed as:
\begin{equation}
	R_{\text{s},o}(k) = \log_2(\text{det}\left(\mathbf{I}_{N_s} + \text{SINR}_{o}(k) \right)).
\end{equation}
Therefore, the total sensing rate $R_{\text{s-sum}}$ across all sensing objects and subcarriers is defined as:
\begin{equation}
	R_{\text{s-sum}} = \sum_{k \in \mathcal{K}} \sum_{o \in \mathcal{O}} \log_2(\text{det}\left(\mathbf{I}_{N_s} + \text{SINR}_o(k) \right)),
	\label{eq:R_sense_sum}
\end{equation}

\subsection{NYU Subcarrier Channel Model}

A framework for creating realistic frequency-selective MIMO channels is offered by the NYU mmWave Channel Model. Based on the number of multipath components $L$ the NYUSIM tool may build each user's and target's subcarrier channel matrix $\mathbf{H}_i(k)$ and $\mathbf{H}_t(k)$, respectively. The $k$th subcarrier channel matrix is supposed to be modeled by the elements of the USA antenna as \cite{1}:

\begin{equation}
	\mathbf{H}(k) = \sum_{\ell=1}^{L} \gamma_{\ell}  e^{j\Phi} e^{-j 2\pi f \tau_{\ell}} \mathbf{a}_r(\phi_{\ell}^r,\theta_{\ell}^r) \mathbf{a}_t^H(\phi_{\ell}^t,\theta_{\ell}^t)\label{array steering vector},
\end{equation}
where \( L \) is the number of multipath components, \( \gamma_{\ell} \) denotes the complex gain of the \( \ell \)-th path, \( f \), $\Phi$ and \( \tau_{\ell} \) are the carrier frequency, the phase shift and the delay associated with the \( \ell \)-th path, respectively, as well as \( \mathbf{a}_r(\phi_{\ell}^r,\theta_{\ell}^r) \) and \( \mathbf{a}_t(\phi_{\ell}^t,\theta_{\ell}^t)\) represent the receive and transmit array response vectors, respectively, where \((\phi_{\ell}^r,\theta_{\ell}^r)\) and \((\phi_{\ell}^t,\theta_{\ell}^t) \) represent the azimuth and elevation AoA at the user and angle of departure (AoD) at the BS for the \( \ell \)-th path, respectively, where \( L=1 \) and \( L > 1 \) for line-of-sight (LOS) and NLOS links scenarios. In LOS scenarios, the channel is dominated by a direct path between the BS and receiver, leading to lower path loss and substantial received power. Thus, the array response vectors associated with the \( l \)th multipath link in Eq. (\ref{array steering vector}) can be modeled in \cite{12}. The blocked user's channel $\mathbf{H}_{i,b}(k) $ is modeled as $\mathbf{H}_{i,b}(k)=\beta\mathbf{H}_i(k)$, where $\beta$ denotes the blockage loss coefficient and usually considered as 20 [dB] for blockage loss owing to human body~\cite{13}. 

\subsection{Problem Formulation}

The goal is to jointly optimize the NOMA users and target probing signals' powers, analog and digital beamforming matrices to maximize the overall communication performance, while ensuring sufficient sensing fidelity for detecting blockage and tracking the target. Hence. the joint optimization of  $R_{\text{s-sum}}$ sensing sum rate and  $R_{\text{c-sum}}$ communication sum rate is formulated as in~(\ref{eq:ISAC_Opt_NOMA}), where $R_{\text{c,s-sum}} = R_{\text{c-sum}} + R_{\text{s-sum}}$ and constraints C1 to C7 mapped to key system requirements for quality of communication service, sensing fidelity, NOMA ordering, HBF, and hardware constraints, respectively. Notably, $\operatorname{tr}\left(\cdot\right)$ denotes trace of a matrix.

\begin{equation}
	\begin{aligned}
		\max_{\substack{
				\mathbf{F}_{\text{RF},{}c,s},\, \mathbf{F}_{\text{BB}{c,s}}(k),\, \mathbf{F}_s(k),\\ 
				\{p_i(k)\}}} \quad & 
		R_{\text{c,s-sum}} \\
		\text{s.t.} \quad 
		& \text{C1: } R_i(k) \geq R_i^{\text{min}}, \quad \forall i,k \\
		& \text{C2: } R_{\text{s},o}(k) \geq R_o^{\text{min}}, \quad \forall o,k \\
		& \text{C3: } p_{\text{weak}}(k) > p_{\text{strong}}(k),  \\
		& \quad \sum_{i=1}^{I} p_i(k) \leq P_k, \quad p_i(k) \geq 0, \quad \forall k \\
		& \text{C4: }\text{tr}(\mathbf{F}_{\text{RF},{c,s}} \mathbf{F}_{\text{BB},{c,s}}(k)) \leq P_{\text{max}}, \forall k \\
		& \text{C5: } \text{tr}(\mathbf{F}_{\text{RF},{c,s}}) = NN_s,
		\text{tr}(\mathbf{W}_i) = N_s, \\ 
		& \quad ~~\text{tr}(\mathbf{F}_{\text{BB},{c,s}}) = NN_s \\
		& \text{C6: } |\mathbf{F}_{\text{RF},{c,s}}(i,j)|^2 = \frac{1}{N_t}, \quad \forall i,j \\
		& \text{C7: } |\mathbf{W}(i,j)|^2 = \frac{1}{N_r}, \quad \forall i,j
	\end{aligned}
	\label{eq:ISAC_Opt_NOMA}
\end{equation}

\section{Proposed Solution: Alternating Optimization Framework for Block Aware ISAC-MC-HBF-NOMA}

This section presents the key components of the proposed solution to the optimization problem in~(\ref{eq:ISAC_Opt_NOMA}). The design integrates joint MC-HBF, dynamic power control, and blockage-aware link switching, tailored for the ISAC-NOMA system. Since the original optimization problem is non-convex, we relax it into a convex form and adopt a round-robin approach. First, the ISAC precoders and combiners are designed under a fixed power allocation. Then, with the optimized precoders and combiners fixed, the power factors are dynamically updated to satisfy the desired quality of service (QoS) requirements. The approach involves the blockage aware mechanism that alert the system to respond to the blockage condition of the ISAC network. These solutions are presented in the following subsections. 

Meanwhile,
\vspace{2pt}
Let $\mathbf{H}_{c,1}(k) \in \mathbb{C}^{N_r \times N_t}$, $\mathbf{H}_{c,2}(k) \in \mathbb{C}^{N_r \times N_t}$ and $\mathbf{H}_t(k)  \in \mathbb{C}^{N_R \times N_t}$ denote the main channel matrices of the NOMA strong user, weak user and dedicated target's channels on subcarrier $k$, respectively. For rank sufficiency, we ensure that $N_R$ Radar antennas at the BS equals $N_s$ superposed sensing symbol streams. A composite sensing-communication stacked channel $\mathbf{H}_{\text{c,s}}(k) \in \mathbb{C}^{N_t \times NN_s}$ per subcarrier is defined as
\begin{equation}
	\mathbf{H}_{\text{c,s}}(k) = \left[ \mathbf{H}_t^\mathsf{T}(k), \mathbf{H}_{c,1}^\mathsf{T}(k) \right]^\mathsf{T}
	\label{eq:stack channel}
\end{equation}
for $N=2$ clusters in this study.
\vspace{4pt}

\subsection{Analog Combiner Design for Sensing and Communication}

Different from singular value decomposition (SVD)-based analog combiner design, we exploit a zero forcing (ZF)-based approach to extract the phase of the channel components. Hence, the analog combiner for each communication user is optimized by extracting the phase structure from the pseudo-inverse of each user's strongest subcarrier channel. Specifically, for user $i \in \{1, 2\}$, exploiting strongest subcarrier $k$, the combiner is given by
\begin{equation}
	\mathbf{W}_i = \frac{1}{\sqrt{N_r}} \exp\left(j\angle\left[\operatorname{pinv}\left(\mathbf{H}_i(k)^\mathsf{H}\right)\right]_{:,1:N_r^{\text{RF}}}\right).
\end{equation}
This formulation supports hardware and combining power constraints and maintains analog phase-only combiner realizations. The same method is deployed by the BS to optimize the sensing receiver combiner $\mathbf{W}_{\text{s}}$, exploiting $\mathbf{H}_t(k)$.

\vspace{4pt}
\subsection{ISAC Analog Precoder Design}

Different from existing work where the analog precoder is optimized from the conjugate phase of the product of combiner matrix and channel matrix, this study optimized ISAC analog precoder $\mathbf{F}_{\text{RF},c,s} \in \mathbb{C}^{N_t \times NN_s}$ from the conjugate negative phase of the composite sensing-communication stacked channel in (\ref{eq:stack channel}), written as:
\begin{equation}
	\mathbf{F}_{\text{RF},c,s} = \frac{1}{\sqrt{N_t}} \exp\left(j\angle\left[-\mathbf{H}_{c,s}(k)^\mathsf{H}\right]\right).
\end{equation}
Structurally, ISAC analog precoder is expressed as 
\begin{equation}
	\mathbf{F}_{\text{RF},c,s} = 	\left[\mathbf{F}_{\text{RF},s},\mathbf{F}_{\text{RF},c}\right].
\end{equation}
This design ensures maximal directional alignment toward the strong user, which also boosts that of weak user for efficient communication, thereby aiding in robust channel estimation and improving ISAC functionality.

\vspace{4pt}
\subsection{ISAC Digital Precoder Design}

The digital baseband precoder $\mathbf{F}_{\text{BB}}$ for communication is computed as a normalized ZF solution based on the effective analog channel $\tilde{\mathbf{H}}_{c,s}\in \mathbb{C}^{N_r \times NN_s}$ after analog processing:
\begin{equation}
	\tilde{\mathbf{H}}_{c,s}(k) = \left[(\mathbf{W}_1^\mathsf{H} \mathbf{H}_{c,1}(k) \mathbf{F}_{\text{c,s}})^\mathsf{T}, (\mathbf{W}_t^\mathsf{H} \mathbf{H}_t(k) \mathbf{F}_{\text{c,s}})^\mathsf{T} \right]^\mathsf{T}.
\end{equation}
Therefore, ZF is computed, using: 
\begin{equation}
	\mathbf{F}_{\text{BB}}(k) = \tilde{\mathbf{H}}_{c,s}(k)^\dagger \cdot \left( \operatorname{tr}\left(\tilde{\mathbf{H}}_{c,s}(k)^\dagger \tilde{\mathbf{H}}_{c,s}(k)^{\dagger \mathsf{H}} \right) \right)^{-\frac{1}{2}}.
\end{equation}
Structurally, ISAC digital precoder $\mathbf{F}_{\text{BB,c,s}} \in \mathbb{C}^{NN_s \times NN_s}$ is expressed as 
\begin{equation}
	\mathbf{F}_{\text{BB}c,s}(k) = \left[\mathbf{F}_{\text{BB},s}(k),\mathbf{F}_{\text{BB},c}(k)\right].
\end{equation}

\vspace{4pt}
\subsection{Dynamic Power Allocation Framework }
After the optimization of the ISAC hybrid precoder and combiners, the optimization problem is reformulated as:

\begin{equation}
	\begin{aligned}
		\max_{\alpha_1,\alpha_2, \alpha_t} \quad & R_{\text{c,s-sum}} \\
		\text{s.t.} \quad & R_2 \geq R_{\text{min}} \\
		& \alpha_1 + \alpha_2 = 0.7, \quad \alpha_1 < \alpha_2, \quad \alpha_1 \geq 0.05,
	\end{aligned}
	\label{eq:optimization}
\end{equation}
where $R_{\text{min}}$ is the weak user's target rate. The power coefficient for dedicated-target is fixed at 30\% of the total power, based on the fact that it enjoys orthogonal access to system resources and could yield optimal performance at any given power factor $\alpha_t \leq 0.3$. Therefore, the optimization problem is relaxed to optimize NOMA users' powers. $\alpha_2$ is clipped to the feasible range \[
\alpha_2 \gets \min(\alpha_c - 0.05,\ \alpha_2 + \delta), \quad 
\alpha_2 \gets \max(0.15,\ \alpha_2 - \delta),
\]
to ensure fairness for strong user and iteration capped at a maximum count, where \( \delta \) represents a small positive step size employed for dynamic power modification.

\vspace{4pt}
\subsection{Blockage-Aware Channel Switching via Reflection Ratio Thresholding}

Instead of relying solely on signal norm comparisons, we adopt a physically motivated approach using reflected signal power to transmitted signal power ratio threshold $\zeta_\text{thr}$, and computed for each user and target's backscattered signal as:
\begin{equation}
	\zeta_{\text{thr}} = \frac{\text{Object}_{\text{RSP}}}{\text{BS}_{\text{TSP}}},
\end{equation}
where $\text{BS}_{\text{TSP}}$ and $\text{Object}_{\text{TSP}}$ denote transmit signal (waveform) power by the BS and reflected signal power from each user. If $\zeta_{\text{thr}} = \rho_o\beta$, the system declares a blockage and switches to an NLOS link for optimization implementation, i.e.,
\[
\text{If}~\zeta_{\text{thr}} = \beta\rho_o,~\mathbf{H}_{c,1}(k) \leftarrow \mathbf{H}_{c,1}^{\text{NLOS}}(k),
~\mathbf{H}_{c,2}(k) \leftarrow \mathbf{H}_{c,2}^{\text{NLOS}}(k).
\] 
If $\zeta_{\text{thr}} = \rho_o$, the system validates no blockage and maintains its LOS link for ISAC-NOMA.

\section{Results}
\subsection{System Parameters Configuration}
The system parameters for the proposed downlink MC-ISAC-NOMA are configured as $N_t=64$, $N_r=4$, $N_R=N_s$, $N_s=4$, $N=2$, $\alpha_t=0.3$, $K=2$, $P_{\text{max}}=1~\text{watt}$, and $\rho=[0.8, 0.5, 0.5]$ for user 1, user 2 and target, respectively. Furthermore, carrier frequency, bandwidth, transmitter power, user 1, user 2, dedicated-target distances to BS, antenna spacing, azimuth and elevation angle (of departure (AoD) same as AoA for LOS link) are configured to 28 GHz, 800 MHz, 30 dBm, 40 m, 120 m, 60 m, 0.5 $\lambda$, $100^o$ and $30^o$ for cluster 1, $140^o$ and $30^o$ for cluster 2 in NYUSIM, respectively. Noise power \(\sigma^2\) is calculated as \(\text{Noise (dB)} = -173 + 10 \log_{10}(\text{bandwidth})\). Moreover, $p_i=\alpha_ip_\text{com}\text{SNR}$ and $p_\text{target}=\alpha_tp_\text{sens}\text{SNR}$. The simulation is carried out under three blockage scenarios: (1 No blockage in LOS, (2 blockage alerted but still maintains LOS, and (3 blockage alerted and switches to deploy unblocked NLOS link) to reveal the performance of the proposed solutions in those ISAC environments. The blocker-aware MC-HBF-ISAC-NOMA system is simulated for a diverse range of SNR values from 0 to 30 dB. In this experiment, the system ensures that the weak user (user 2) consistently achieves a minimum communication rate of 2 to 4 bps/Hz under dynamically adjusted power coefficients. Two blockage profiles are considered to demonstrate the system’s adaptability.

\subsection{Overall Achievable MC-ISAC-NOMA Sum Rate under Blockage Conditions}

\begin{figure}[!t]
	\centering
	\includegraphics[width=0.9\columnwidth]{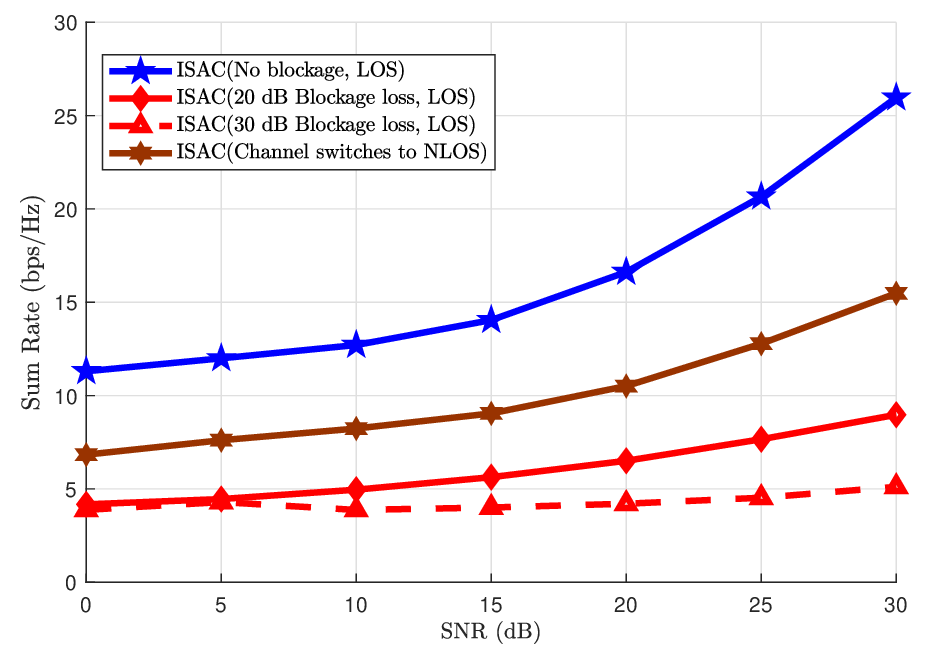}
	
	\caption{Achievable overall sum rates for the proposed ISAC-NOMA system configured for 2 b/s/Hz minimum rate of weak user and tested for two versions of blockage loss coefficients.}\label{Figure3}
\end{figure}

Figure~\ref{Figure3} presents the overall achievable ISAC-NOMA system's sum rate performances, deploying a joint waveform MC-HBF and dynamic power optimization configured for 2 bps/Hz weak user's minimum rate. The results show that the sum rate grows with increasing SNR for all scenarios, reflecting the natural capacity improvement associated with higher transmit power. Particularly, at 15 dB SNR, the 20 dB and 30 dB blockage loss coefficients lead to the 57\% and 71\% sum rate degradation compared to the unblocked condition. These degradations are minimized to 36\%, exploiting intelligent channel switching and dynamic power reallocation.

\subsection{Disaggregated Sensing and Communication Rates}
\begin{figure}[!t]
	\centering
	\includegraphics[width=0.9\columnwidth]{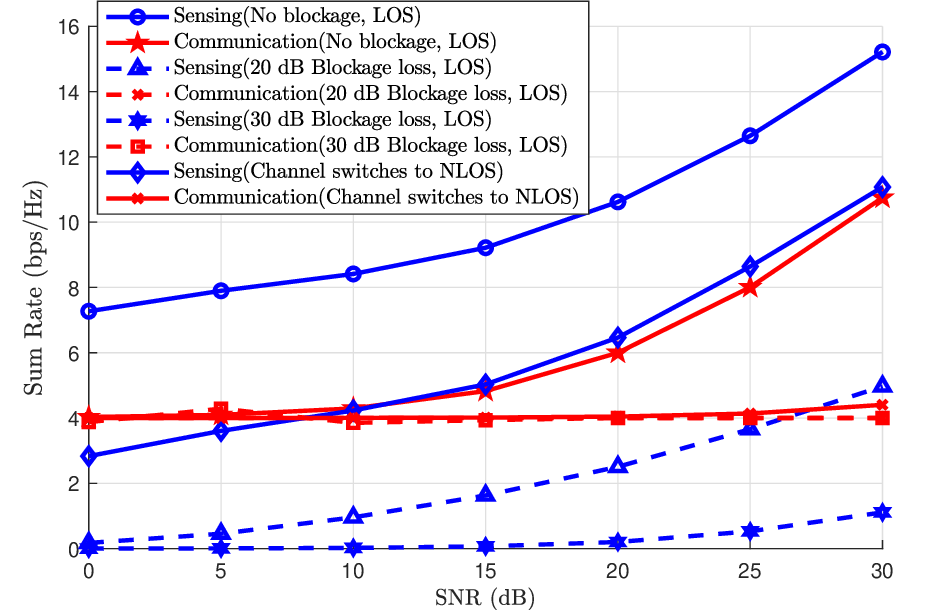}
	
	\caption{Achievable sum-rates for both sensing and communication, revealing their composition in the results shown in Fig.~\ref{Figure3}.}\label{Figure4}
\end{figure}

In Figure~\ref{Figure4}, the sum rate is disaggregated into communication and sensing components, providing further insight into the contributions of each domain to the total performance in Fig.~\ref{Figure3}. During no blockage, the sensing achieves higher sum rate than the communication at all SNR regimes. However, during the three blockage conditions, communication performs better than sensing at 0 to 10 dB SNRs. Furthermore, this dominance of communication performances proceeds until 26 dB SNR compared to that of sensing during two blockage loss conditions without switching to NLOS link, indicating the higher degradation impact of blockage on the strength of backscattered signal during passive sensing compared to the impact on communication. Notably, the communication sum rate has a negligible performance gain during NLOS switching scenario compared to the substantial performance gain of sensing at 10 to 30 dB SNRs. This trend confirms the efficacy of power coefficient adaptation, where excess power at high SNR is reallocated to sensing without significantly compromising communication quality. This performance trade-off highlights the system's ability to adaptively balance spectrum sharing between sensing and communication tasks as environmental conditions evolve.

\subsection{Per-User Achievable Rates under Dynamic Power Allocation}

\begin{figure}[!t]
	\centering
	\includegraphics[width=0.9\columnwidth]{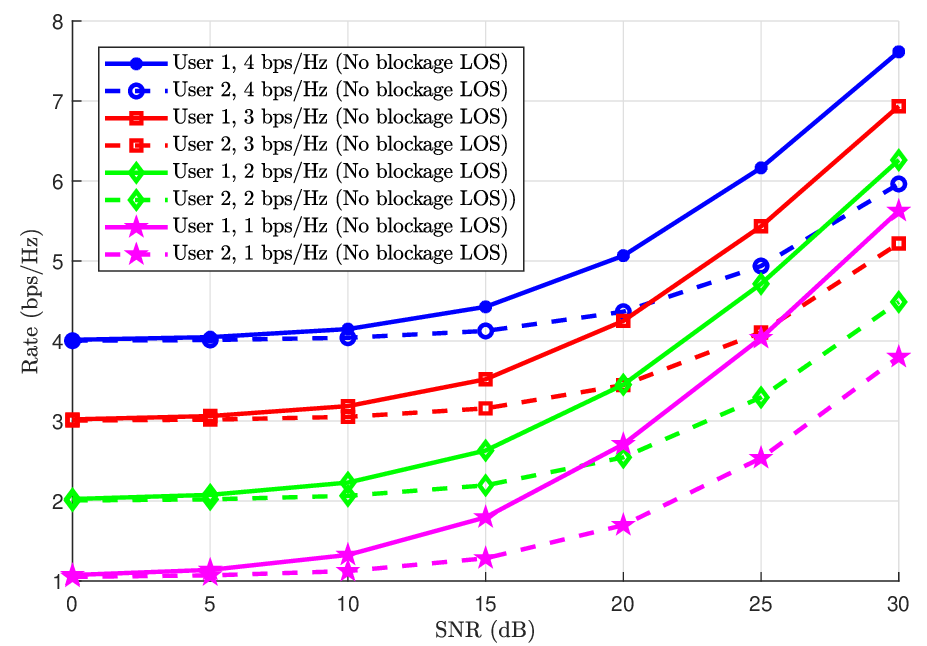}
	
	\caption{Achievable rates for each NOMA user, deploying two subcarriers for transmission and revealing the feasibility of the proposed power scheme for attaining diverse minimum rates targeted for weak users in ISAC network.}\label{Figure5}
\end{figure}

Figure~\ref{Figure5} illustrates the individual rate trends for each user across the two subcarriers, shedding light on how the dynamic power allocation framework maintains fairness while optimizing rate under no blockage condition. The strong user consistently achieves higher rates than the weak user, as expected from the power-domain NOMA structure. More importantly, the weak user's rate curve reaches the minimum desired rate at the lowest SNR and increases progressively with higher SNR values. This performance trend verifies that the adaptive algorithm converges effectively and enforces the minimum QoS constraint.

\section{Conclusion}
This paper presented a blockage-resilient MC-ISAC-NOMA framework that leveraged HBF and dynamic power allocation to overcome link performance degradation caused by physical obstructions. The system employed joint HBF matrix design and a channel-aware switching mechanism to dynamically adapt to changing propagation conditions. Additionally, a rate-constrained power allocation algorithm ensured fairness by maintaining the minimum communication rate for the weak NOMA user. Simulation results demonstrated that the proposed approach significantly enhanced sensing performance and preserved the communication throughput in mid-to-high SNR conditions. Notably, during NLOS switching, sensing rate gains exceeded 400\% at 15 dB SNR, with minimal impact on communication rates. Furthermore, dynamic adaptation guaranteed that the weak user achieved the desired QoS even under severe blockage, reinforcing the system’s convergence efficiency and fairness. Thus, the blocker-aware MC-ISAC-NOMA architecture offered a practical and efficient solution for next-generation wireless networks, enabling robust sensing and communication capabilities in dynamic environments.

\section*{Acknowledgment}
This work was supported by the National Research Foundation of Korea (NRF) grant funded by the Korea government (MSIT) (No. RS-2024-00409492).

\end{document}